\documentclass{elsart}

\usepackage{natbib}
\usepackage{graphicx}

\usepackage{amssymb}
\usepackage{amsmath}

\def\url#1{{\ttfamily\def\/{/\discretionary{}{}{}}#1}}
\def\bibcode#1{}

\begin{document}

\begin{frontmatter}
\title{Sunyaev-Zel'dovich polarization simulation}
\author[address1]{Alexandre Amblard\thanksref{aaemail}},
\author[address1,address2]{Martin White\thanksref{mwemail}}
\address[address1]{Department of Astronomy, University of California,
Berkeley, CA, 94720}
\address[address2]{Department of Physics, University of California,
Berkeley, CA, 94720}
\thanks[aaemail]{E-mail: amblard@astro.berkeley.edu}
\thanks[mwemail]{E-mail: mwhite@astro.berkeley.edu}

\begin{abstract}
Compton scattering of Cosmic Microwave Background (CMB) photons on
galaxy cluster electrons produces a linear polarization, which
contains some information on the local quadrupole at the cluster
location. We use N-body simulations to create, for the first time,
maps of this polarization signal. We then look at the different
properties of the polarization with respect to the cluster position
and redshift.
\end{abstract}

\end{frontmatter}

\section{Introduction}
Photons from the last scattering surface (the cosmic microwave
background; CMB) propagate toward us, interacting with the matter in
between. These interactions cause a change in the temperature and
polarization pattern of the CMB. For instance, Compton scattering can
produce linear polarization if the interacting medium is illuminated
by a quadrupolar radiation field. \\\cite{SZ80} first explored the
different effects on CMB polarization that galaxy clusters, here the
interacting medium, could produce and distinguished 3 sources : the
primordial CMB quadrupole seen by the clusters, the quadrupole
produced by a first interaction inside the cluster, and the transverse
velocity of the cluster. Similar effects have been proposed more
recently \citep{CM02,D03}. Though these are very small signals, it
has been advocated that one can get interesting information on the CMB
quadrupole \citep{KL97,P04} and the cluster transverse velocity
\citep{SZ80,SS99,AS99} through these effects. \cite{CB03} have shown 
that the contribution from the primordial quadrupole dominates the
signal, it will therefore be the focus of our work.\\ In this paper we
present the first map of the polarization arising from the primordial
quadrupolar CMB anisotropies (though see \cite{D02} and
\cite{M04} for simulations of other SZ effects) and look for the
typical properties of the signal. We describe in the following our
physical model of the induced polarization, we then show how we
simulate the effect under reasonable assumptions, we finish by showing
interesting properties of this simulation.

\section{Model}
\label{par:model}

The primordial CMB quadrupole is generated by two effects : the SW
(Sachs-Wolfe) effect and the ISW (Integrated SW) effect which for
adiabatic fluctuations in the cluster reference frame can be written
as \citep{SW67} :

\begin{eqnarray}
\hspace{3cm}\frac{\Delta T_{\mathrm{SW}}(\hat n)}{T} &=& - {1\over3}\; {\Phi(\hat n,z_{\rm CMB})} \\
\hspace{3cm}\frac{\Delta T_{\mathrm{ISW}}(\hat n)}{T} &=& - 2\int_{z_{\rm CMB}}^{z_{\rm clus}}\dot{\Phi}({\hat n},z)dz
\end{eqnarray}

where $z_{\rm clus}$ is the cluster redshift, $z_{\rm CMB}\simeq10^3$
is the redshift of the last scattering surface, and $\hat n$ is the
angular position of the cluster. The cross section for Compton
scattering is :

\begin{equation}
\frac{d\sigma}{d\Omega}=\frac{3\sigma_T}{8\pi}|{\hat\epsilon}_{\rm in}\cdot{\hat\epsilon}_{\rm out}|^2
\end{equation}

where ${\hat\epsilon}_{\rm in}$, ${\hat\epsilon}_{\rm out}$ stand
respectively for the input and output photon polarization vector, and
$\sigma_T$ for the Thomson cross-section. Using the stokes parameters
Q and U, defining our coordinate system (centered on the cluster) with
the axis $\hat z$ in the line of sight direction ($x$ and $y$ axis
will define $Q$ and $U$ basis), and integrating on all incoming photon
directions, we get :

\begin{eqnarray}
\hspace{3cm}Q\,({\hat z}) &=& \frac{3\sigma_T}{16\pi}\int{\Delta T(\hat n) \sin^2\theta \cos 2\phi d\Omega}\\
\hspace{3cm}U\,({\hat z}) &=& \frac{3\sigma_T}{16\pi}\int{\Delta T(\hat n) \sin^2\theta \sin 2\phi d\Omega}
\end{eqnarray}

Substituting $\Delta T_{\mathrm{SW}}(\hat n)+\Delta
T_{\mathrm{ISW}}(\hat n)$ for $\Delta T(\hat n)$ and $\tau_{\rm clus}$
for $\sigma_T$ in the above, we obtain the polarization created by
clusters for a given potential $\Phi$.

\section{Simulations}

In order to create a $\Phi$ field with which we compute the local
quadrupole, we generated a cube of $256^3$ points covering 30 Gpc in
size at $z=0$, which gave us enough space to trace back to the last
scattering surface ($z\simeq 1100$) with sufficient resolution (around
100 Mpc/h) to compute the integrals in \textsection\ref{par:model}.\\
We used a simple Harrison-Zel'dovich, or scale-invariant, spectrum to
model the potential fluctuations as we only look at very large scales.
We then used an N-body simulation (see \cite{V04,A04} for more details) to
compute $\tau_{\rm clus}$ in different $z$ slices (35 from z = 0 to
2). The last step is to combine all of these ingredients :\\
\begin{eqnarray}
Q(\hat u) = & -\sum_{i}\frac{3\tau_{\rm clus}(z_i)}{16\pi}\sum_{\theta,\phi}
\left\{{\Phi(r_i\hat u + r^{\scriptscriptstyle{\rm CMB}}_i\hat n,z_{\rm \scriptscriptstyle{CMB}})\over 3} + 2 \sum_j \dot{\Phi}(r_i\hat u + r_j\hat n, z_j)\right\}\nonumber\\
&\times  \sin^2\theta \cos2\phi \Delta z\Delta\Omega 
\end{eqnarray}

$\tau_{\rm clus}(z_i)$ stands for the optical depth at redshift $z_i$
(corresponding to one of our N-body simulation slices), $r_i$
represents the distance to this redshift, $\hat n$ is the unit vector
of direction $(\theta,\phi)$ (we map the directions using the
HEALPix\footnote{http://www.eso.org/science/healpix} package), and
$r_i^{\rm CMB}$ is the distance to the last scattering surface from
the redshift $z_i$. We obtain $U(\hat u)$ by substituting $\cos2\phi$
by $\sin2\phi$. \\ From these Q and U maps we can compute the
polarization fraction and angle or the E and B maps. In order to keep
our algorithm simple and fast we do not use any interpolation on our
$256^3$ map, but we test by changing the resolution of the different
quantities that our accuracy is better than 10\%. We note that our
result includes only contributions to the signal from clusters with
$z<2$. Though these clusters are the dominant signal for our purposes,
source at higher redshift may contribute significantly to the signal
on large scales.

\section{Results}
\begin{figure}[h!]
\begin{center}
\begin{tabular}{ll}
\includegraphics[width=7cm]{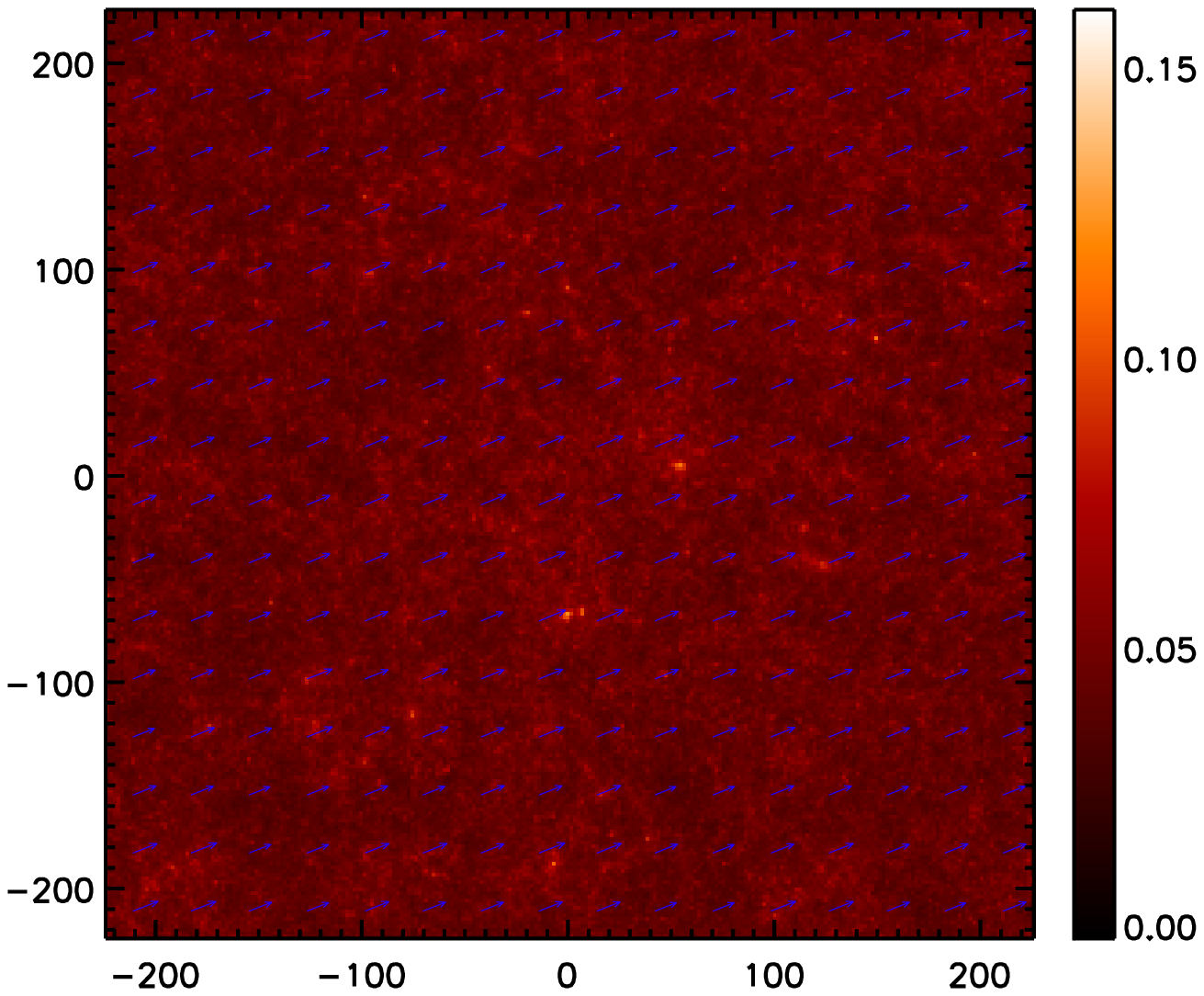}&
\includegraphics[width=7cm]{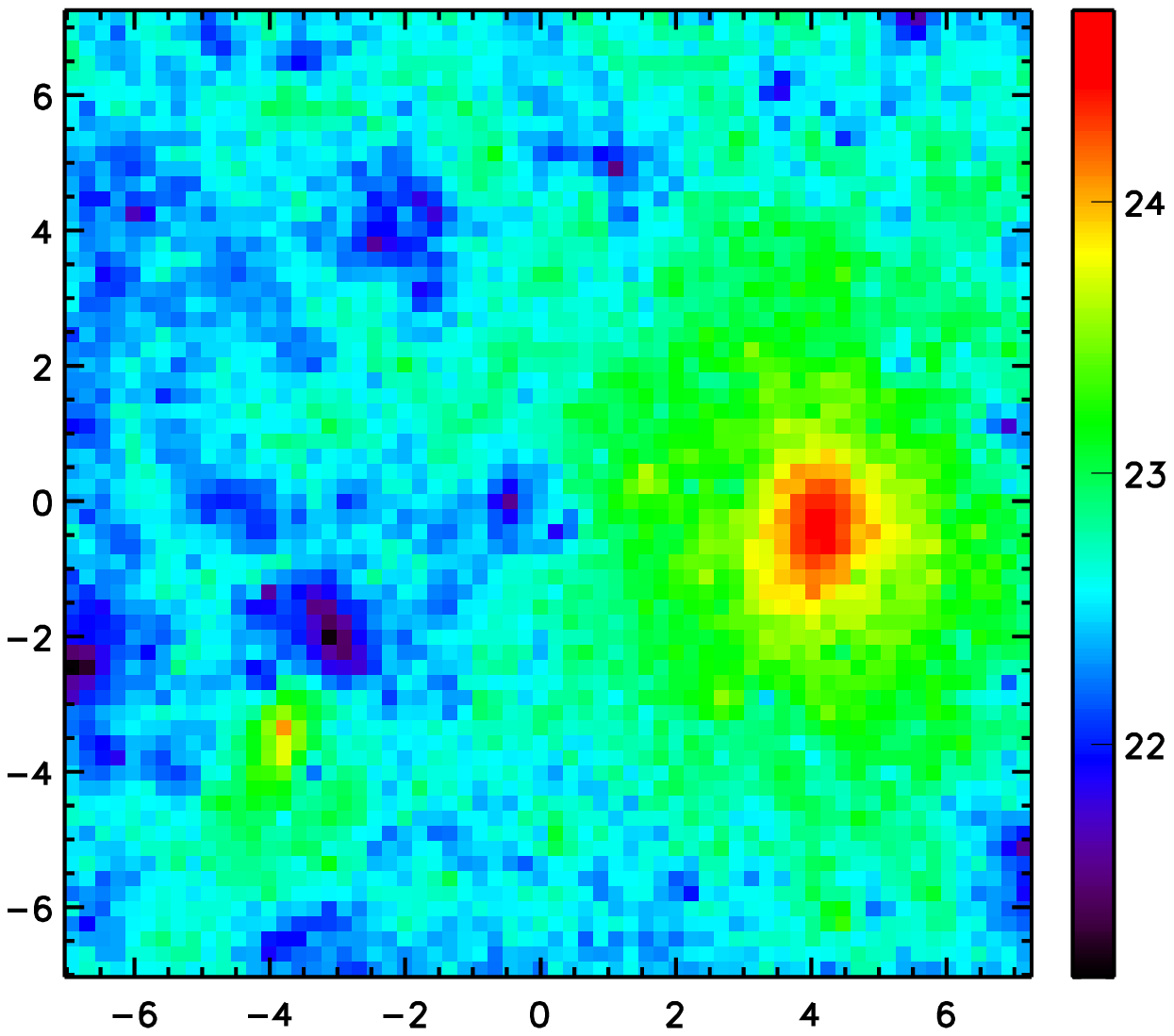}\\
\includegraphics[width=7cm]{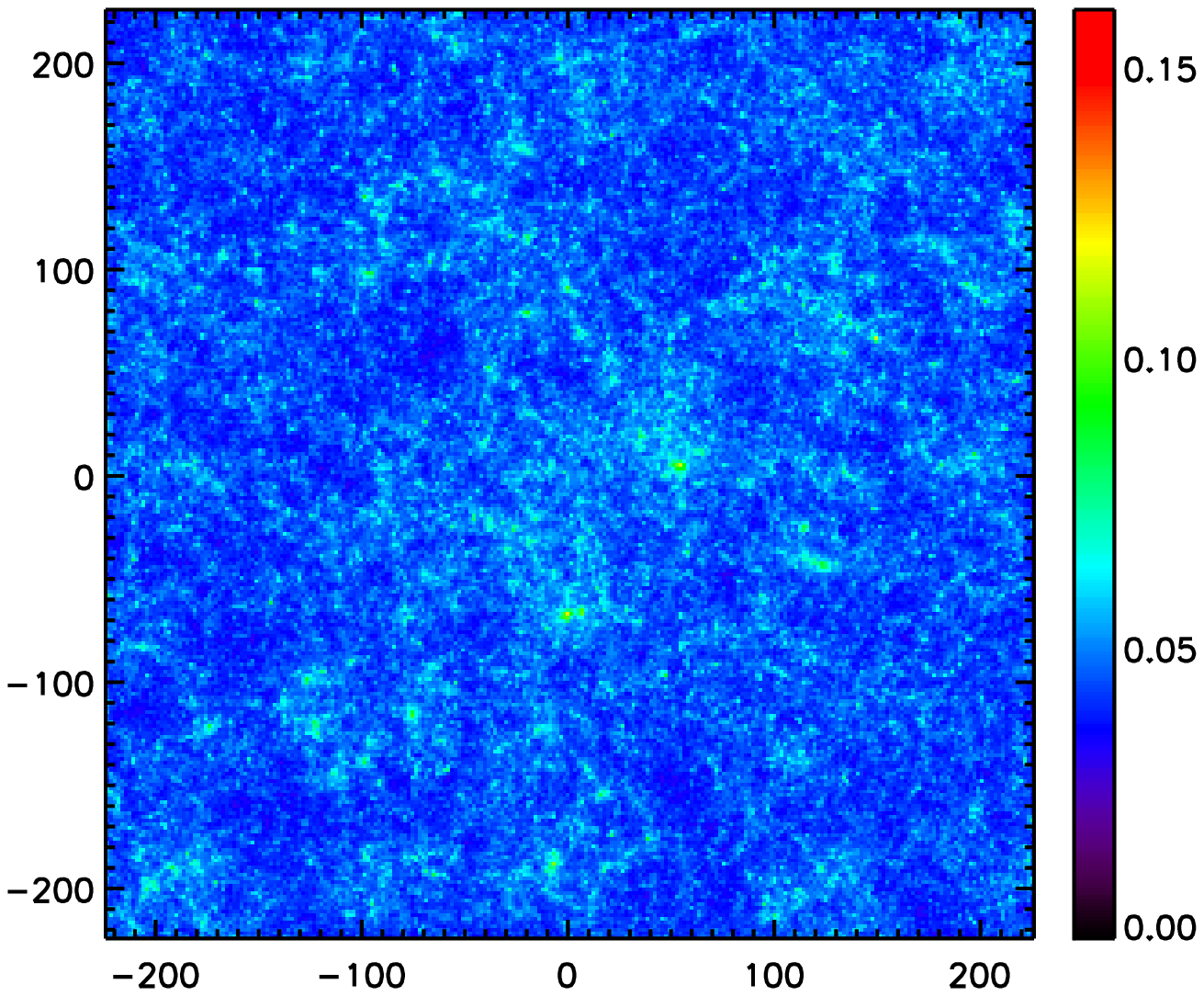}&
\includegraphics[width=7cm]{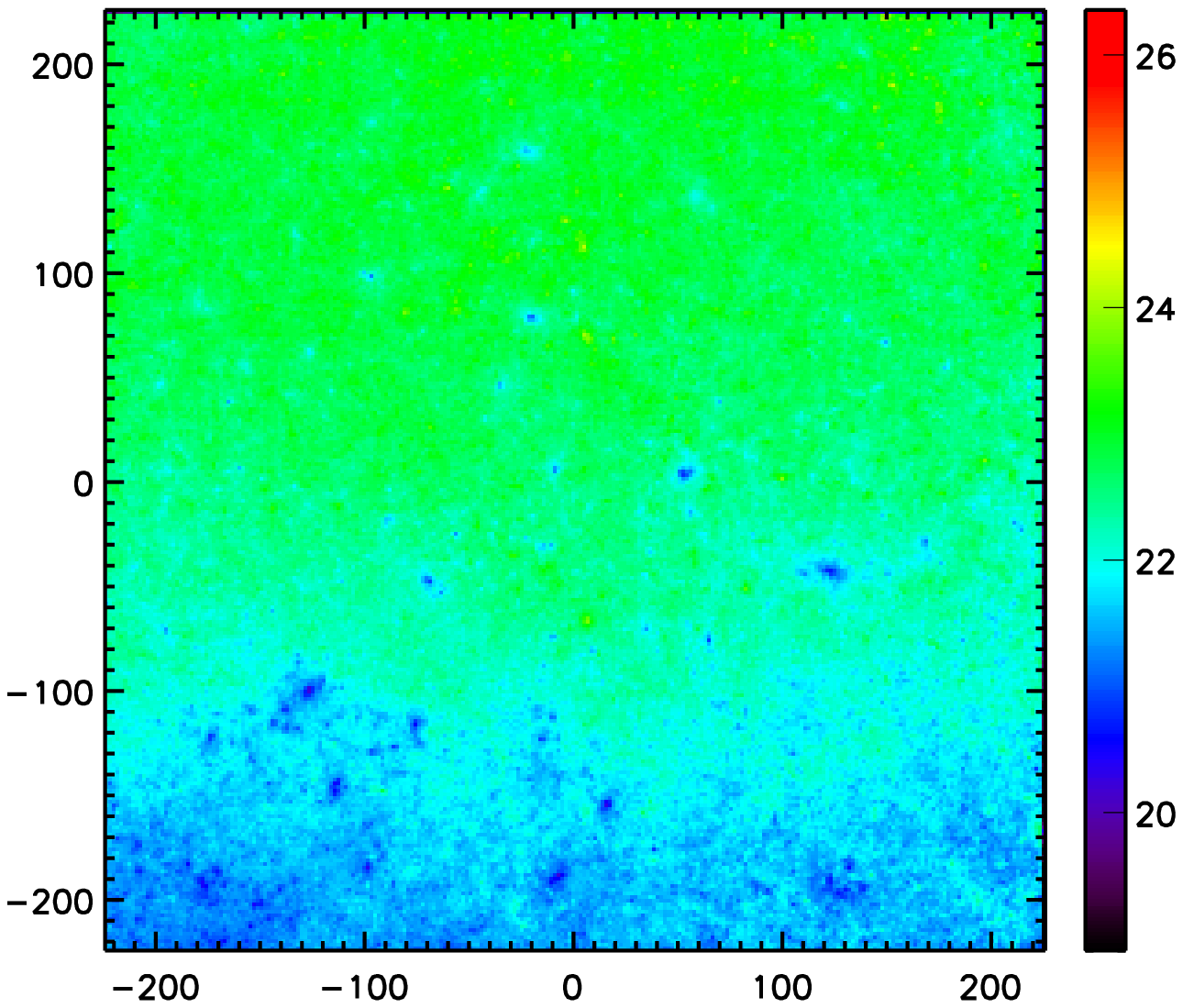}\\
\end{tabular}
\end{center}
\caption{Top Left : Polarization pattern of 7.5$\times$7.5 degrees 
(axis scales in arcminutes) patch located around (0,0) : the color
scale (maximum set to 0.16 $\mu$K in order to reveal more structures)
indicates the polarization amplitude $P=\sqrt{Q^2+U^2}$, arrows
indicate the polarization angle. Top Right : Polarization angle (in
degrees) for the area around (2,-67) on the left plot, this area
contains 2 massive clusters of 1.15$\times10^{15}\mathrm{M}_\odot$
(red spot on the right) and 6.25$\times10^{14}\mathrm{M}_\odot$ (blue
spot on the left) at a respective redshift of 0.25 and 1.11. Bottom :
polarization amplitude (left) and angle (right in degrees) of our
7.5$^\circ$ patch, notice the 2 effects on the polarization angle
pattern, one due to the alignment with the quadrupole (large scale
gradient), and one due to the different redshift values of the
clusters (small scales).}
\label{fig:bigpol}
\end{figure}

\begin{figure}[h]
\begin{center}
\includegraphics[width=14cm]{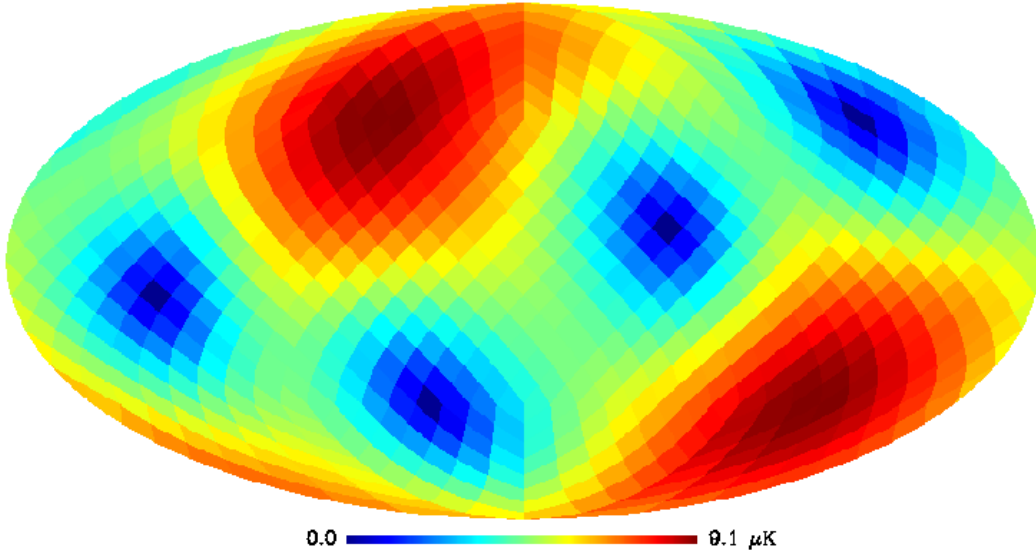}
\end{center}
\caption{Full sky pattern (produced using the HEALPix package) of 
P/$\tau$ for redshift $z=0.034$ (our first slice). The 4 minima
correspond to the 4 minima at 45 degrees of the hot and cold spot of
the local CMB quadrupole, the 2 maxima correspond to the ones at 90
degrees.}
\label{fig:bigpol2}
\end{figure}

\begin{figure}[h]
\begin{center}
\begin{tabular}{ll}
\includegraphics[width=6.5cm]{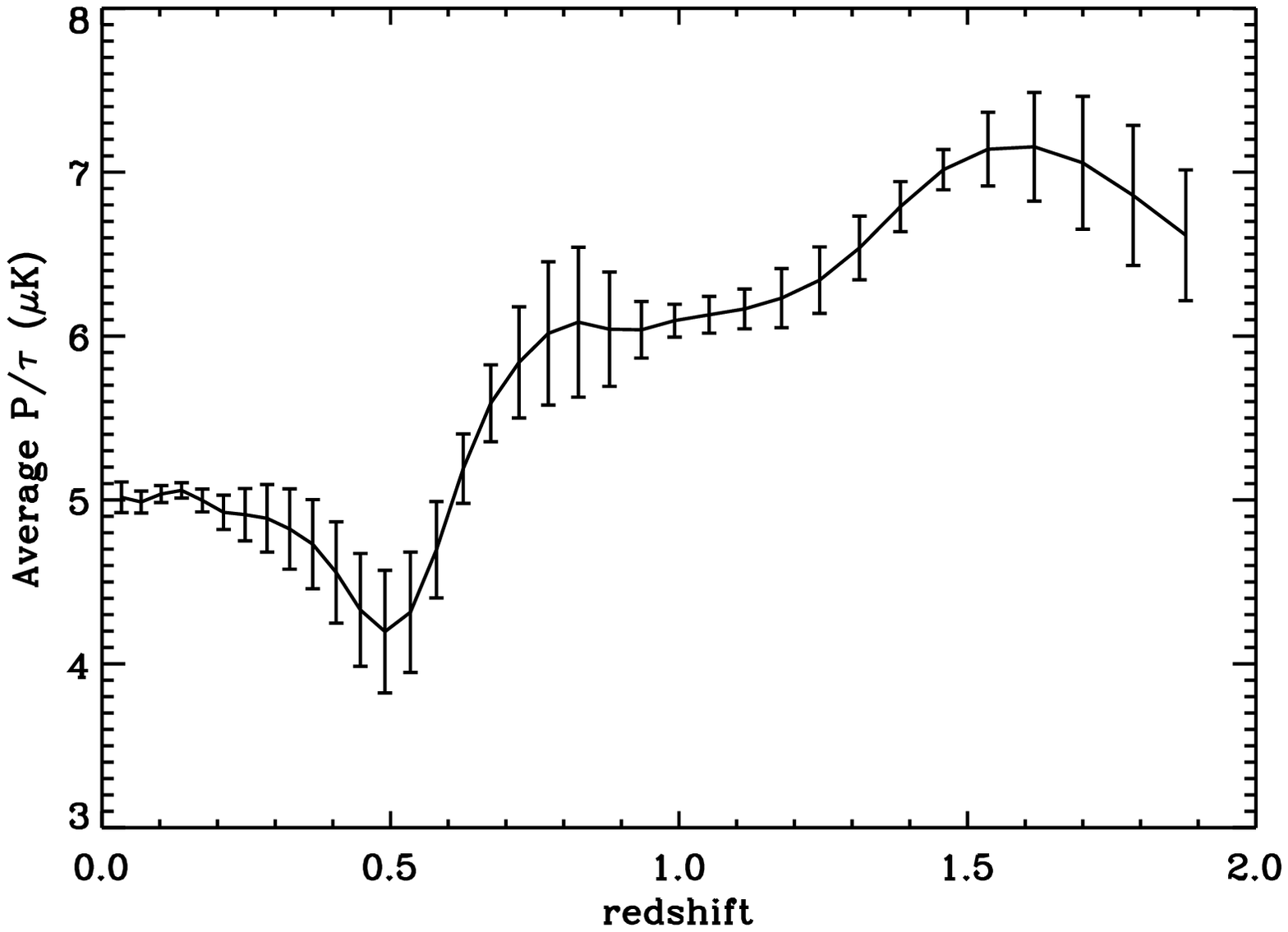}&
\includegraphics[width=7.2cm]{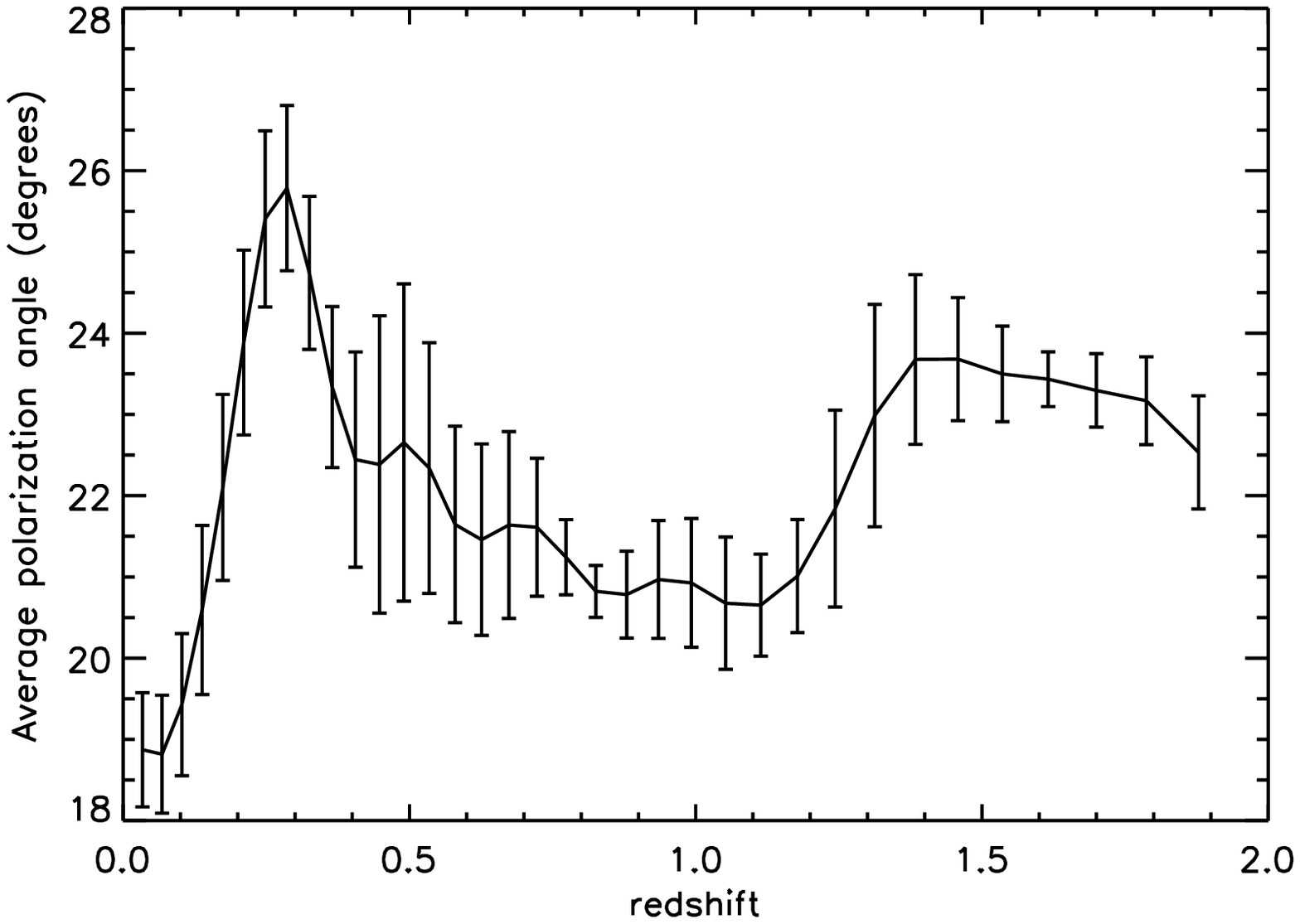}
\end{tabular}
\end{center}
\caption{Average variation (over our 7.5$\times$7.5 degrees patch) 
of the $P/\tau$ (left) and of the polarization angle (right) versus
redshift, error bars represent the 1 $\sigma$ dispersion of these
values.}
\label{fig:varvsz}
\end{figure}

Our basic result is shown in Figure \ref{fig:bigpol}, which displays
the polarization pattern expected over a 7.5$^\circ$ patch of the
sky. The polarization amplitude goes up to 0.16 $\mu$K, the maximum
amplitude depends of the cluster position (here maximum around
$(l,b)\simeq(40,60)$ or $(-40,240)$, see figure \ref{fig:bigpol2}) and
redshift (here maximum achieved around $z \simeq 1.6$, see figure
\ref{fig:varvsz}) with respect to the value of the quadrupole. 

\begin{figure}[h]
\begin{center}
\begin{tabular}{ll}
\includegraphics[width=6.5cm]{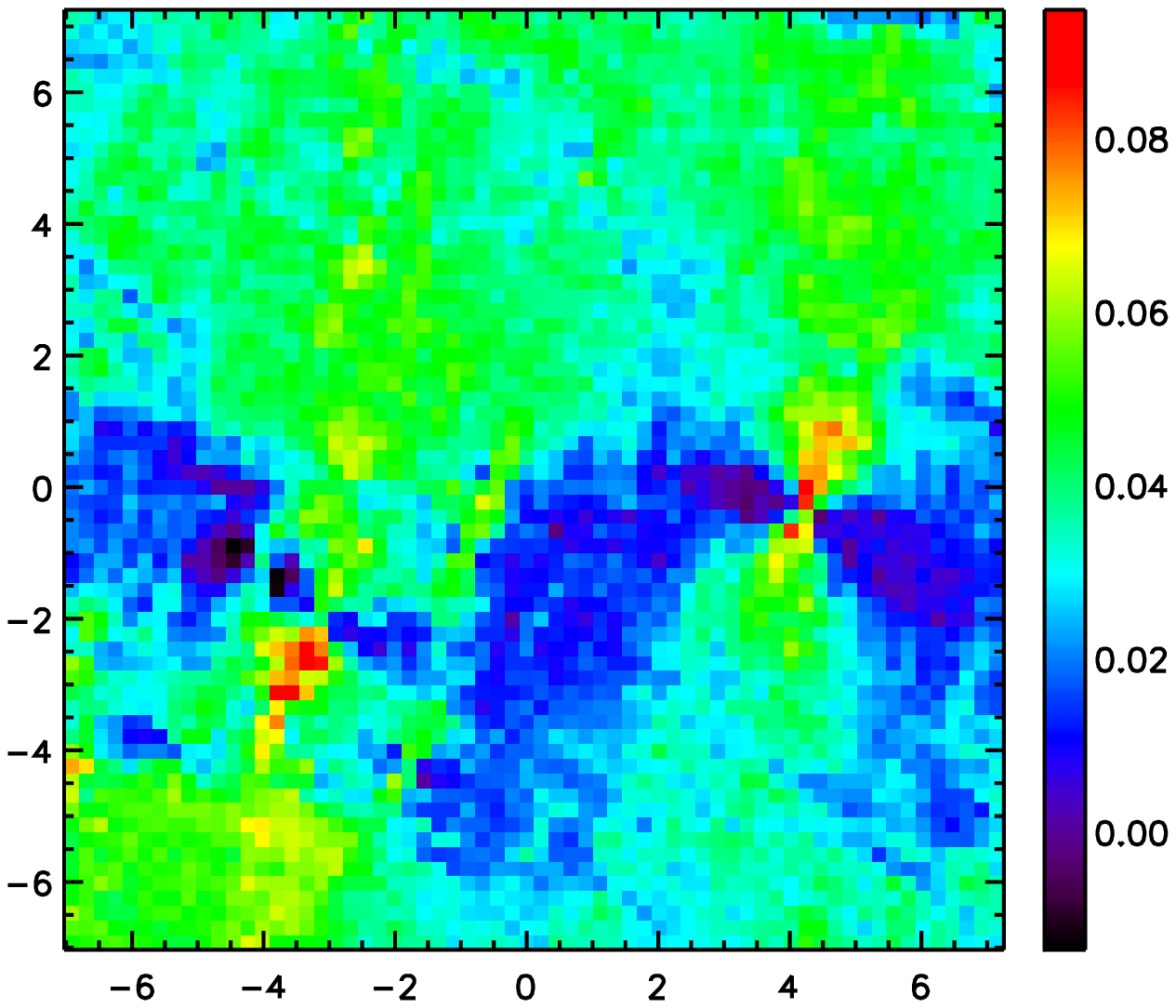}&
\includegraphics[width=6.5cm]{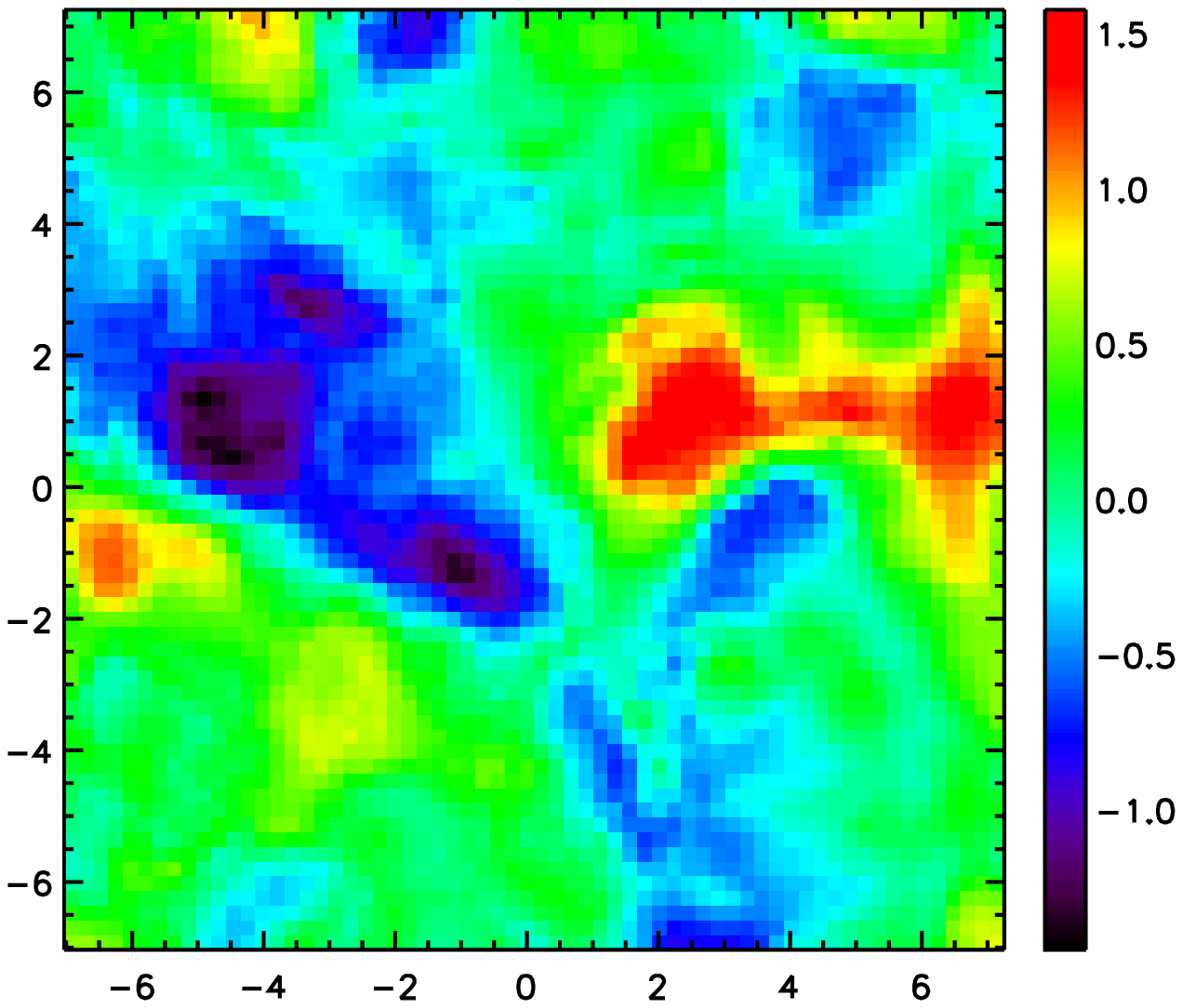}
\end{tabular}
\end{center}
\caption{Part of the 7.5$\times$7.5 degree map centered around 2 massive
clusters (same area as Figure \ref{fig:bigpol} top right). The left map
represents the polarized SZ B mode, the right map the contribution of
the lensing to the B mode. The amplitutde of the polarized SZ ($\simeq
0.1 \mu\mathrm{K}$) is about 30 times smaller than the lensing of the CMB
($\simeq 3 \mu\mathrm{K}$).}
\label{fig:cluszoom}
\end{figure}

\newpage
Our particular realization does not maximize the polarized signal
coming from the SZ, but gives a reasonable estimate of its amplitude
which in any case would not exceed a few tenths of 1$\mu$K.  The
signal for polarized Sunyaev-Zel'dovich effect is therefore not
detectable with present technology
(eq. B2K2\footnote{http://www.astro.caltech.edu/\~{}lgg/boomerang\_front.htm}
and Maxipol\footnote{http://groups.physics.umn.edu/cosmology/maxipol/}
should achieved around 50 $\mu$K/arcmin) on any scales. Future surveys
(Polarbear II\footnote{http://bolo.berkeley.edu/polarbear}, CMBPol),
which should achieve a sensitivity around 1$\mu$K/arcmin, may have
enough sensitivity to measure the effect on the larger scales
($\ell<200$). On the figure \ref{fig:cluszoom}, we compared the
polarized CMB lensing effect with the polarized SZ by showing a patch
of our simulated sky around the 2 massive clusters in figure
\ref{fig:bigpol} (red spot on the right and blue spot on the left with 
a respective mass of 1.15$\times10^{15}\mathrm{M}_\odot$ and
6.25$\times10^{14}\mathrm{M}_\odot$ and a respective redshift of 0.25
and 1.11). The B mode amplitude from the polarized SZ is around 0.1
$\mu$K, much lower than the one coming from the CMB lensing (around
3$\mu$K). Therefore even if the sensitivity to detect the polarized SZ
could be achieved one would have to disentangle it from the CMB
lensing signal. The very different pattern of the two B mode signals
(CMB lensing is much smoother and correlated to the E mode gradient)
could maybe help in such a task.

\begin{figure}[h]
\begin{center}
\includegraphics[width=14cm]{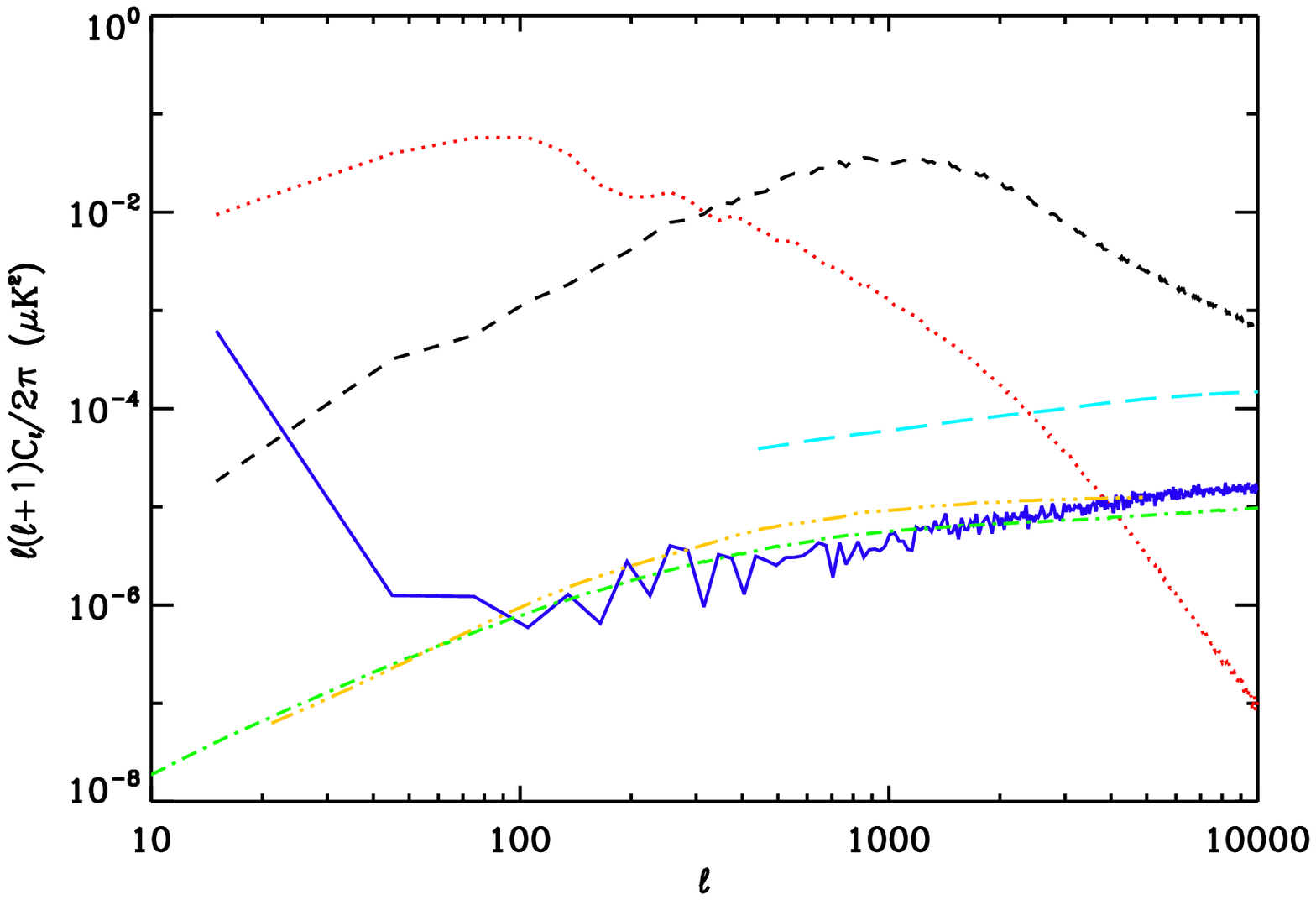}
\end{center}
\caption{The B mode power spectra of the primordial CMB with tensor/scalar 
ratio of 1 (red dots), lensed CMB (black dashes) and Sunyaev
Zel'dovich effect (blue solid line). Previous estimate from \cite{C04}
(green dot dashed line), \cite{H00} (orange dash-triple dotted line)
and recent estimate from \cite{L04} (light blue long dashed
line) are also shown for comparison. The SZ power spectrum can be
subdivided in two parts : on large scale ($\ell<40$) the geometrical
projection effect of the quadrupole dominates, on small scale
($\ell>100$) the variation of the optical depth dominates.}
\label{fig:clcomp}
\end{figure}

From the different maps (polarized SZ, primordial CMB, lensing of the
CMB), we compute the power spectra (see figure \ref{fig:clcomp}). The
B mode of the polarized SZ is very small relative to the lensing of
the CMB and to the primordial CMB (here we took an ``optimistic''
tensor to scalar ratio of unity, $r=1$) from $\ell=20$ to $10000$.
Only the B mode spectra are shown on figure \ref{fig:clcomp} but the E
mode spectrum of the polarized SZ is identical to the B mode ones, so
that polarized SZ E mode are even more subdominant compared to other
sources. Our estimated polarized SZ spectrum has two main features : a
steep decrease on large scales ($\ell$ between $0$ to $50$) and a slow
increase on small scales ($\ell$ between $100$ to $10000$).The
former is produced by the variation of the projected quadrupole onto
the line of sight (see for instance the quadrupole pattern at
$z=0.034$ on figure \ref{fig:bigpol2}). The value of the power on
these scales is however subject to a huge cosmic variance error (the
relative error due to cosmic variance goes basically as
$\sqrt{{2}/{(2\ell+1)/f_{sky}}}$, where $f_{sky}$ represents the
fraction of the sky covered, here 0.13\%), so that our prediction is
to be taken with caution. This rising on large scales could be a
window to measure the polarized SZ if the tensor perturbation were
small enough (typically if $r$ was around $0.01$).

The power on small scales is produced by the anisotropies in the
optical depth $\tau$, and is increasing like $\ell^{0.5}$. The estimated
power spectra by \cite{H00} and \cite{C04} (respectively orange
dash-triple dotted and green dot-dashed lines on figure
\ref{fig:clcomp}) match quite well our estimate on small
scales ($\ell > 100$). That is somewhat surprising as our reionization
redshift is quite low : our simulation does not include structures at
$z>2$ whereas the other estimates go at least to a redshift 5. We
think that this match shows that our estimate have the same order of
magnitude, but including higher $z$ may oncrease the signal further.
The numerical estimate from \cite{L04} is an order of magnitude higher
than the semi-analytical results from \cite{H00} and \cite{C04}, and
from our own estimate. Our low reionization redshift could explain
this discrepancy, the difference with \cite{H00} and \cite{C04} could
lie in the details of their reionization model.\\ Another window
on the polarized SZ signal is the polarization angle. On Figure
\ref{fig:bigpol}, the polarization angle changes very slowly with the
angular position (6.8 degrees in the 7.5 degrees field of view)
especially at low redshift, as it is determined mainly by the angle
between the line of sight and the CMB quadrupole direction. It also
changes quite slowly with respect to the cluster redshift (Figure
\ref{fig:varvsz}), due to the high correlation between quadrupole
orientation at different redshift (the correlation length is about
0.1). It implies that one gains very little by measuring the
quadrupole at different redshifts between 0 and 2 due to this large
correlation (in agreement with \cite{P04}).  Furthermore, 2 clusters
in the same direction but with different polarization angle will in
fact be at different redshifts (like the 2 clusters in figure
\ref{fig:bigpol} at redshift 0.25 and 1.1), this difference in
redshift being greater than the correlation length.

\section{Conclusion}

Compton scattering of CMB photons in the hot intra-cluster medium in
massive halos generates a linear polarization proportional to the
cluster optical depth and the local quadrupole of the CMB intensity.
We have presented the first maps of this effect based on numerical
simulations. With the procedure we have outlined in our simplified
reionization model, the maps are accurate to 10\%. Our simulations
confirmed that the level of polarization due to this effect is rather
small (a few tenth of $\mu$K), though we computed the effect only from
sources at redshift lower than 2. The additional power coming from
higher redshift would probably increase the statistical significance
of the polarized SZ, but it should have little effect locally. We
computed the power spectrum of our map and found similar results to
\cite{H00} and \cite{C04}. The polarization angle variation is dominated 
by the geometrical effect between the quadrupole direction and the
observation direction, its value changes slowly in both angular
position and redshift.

{\bf Acknowledgments:}\newline

We would like to acknowledge the use of the HEALPix package for our
map pixellisation \citep{G99} and thank our anonymous referee for
usefull comments.

\end{document}